\documentclass[twocolumn,nofootinbib]{revtex4}

\input psfig.sty

\usepackage{graphicx}

\begin{document}

\title{Observational constraints on late-time $\Lambda(t)$ cosmology}

\author{S. Carneiro\footnote{ICTP Associate Member}$^{1}$\footnote{saulo.carneiro@pq.cnpq.br},
M. A. Dantas$^2$\footnote{aldinez@on.br}, C.
Pigozzo$^1$\footnote{cpigozzo@ufba.br} and J. S.
Alcaniz$^2$\footnote{alcaniz@on.br}}

\affiliation{ $^1$Instituto de F\'{\i}sica, Universidade Federal da
Bahia, Salvador - BA, 40210-340, Brasil \\ $^2$Departamento de
Astronomia, Observat\'orio Nacional, Rio de Janeiro - RJ, 20921-400,
Brasil}

\date{\today}

\begin{abstract}

The cosmological constant $\Lambda$, i.e., the energy density stored
in the true vacuum state of all existing fields in the Universe, is
the simplest and the most natural possibility to describe the
current cosmic acceleration. However, despite its observational
successes, such a possibility exacerbates the well known $\Lambda$
problem, requiring a natural explanation for its small, but nonzero,
value. In this paper we study cosmological consequences of a
scenario driven by a varying cosmological term, in which the vacuum
energy density decays linearly with the Hubble parameter, $\Lambda
\propto H$. We test the viability of this scenario and study a
possible way to distinguish it from the current standard
cosmological model by using recent observations of type Ia supernova
(Supernova Legacy Survey Collaboration), measurements of the
baryonic acoustic oscillation from the Sloan Digital Sky Survey and
the position of the first peak of the cosmic microwave background
angular spectrum from the three-year Wilkinson Microwave Anisotropy
Probe.

\end{abstract}

\maketitle

\section{Introduction}
The nature of the mechanism behind the current cosmic acceleration
constitutes a major problem nowadays in Cosmology \cite{rev}. Even
though almost all observational data available so far are in good
agreement with the simplest possibility, i.e., a vacuum energy plus
cold dark matter ($\Lambda$CDM) scenario, it is becoming rather
consensual that in order to better understand the nature of the dark
components of matter and energy, one must also consider more complex
models as, for instance, scenarios with interaction in the dark
sector \cite{cq}.

In this regard, the simplest examples of interacting dark
matter/dark energy  models are scenarios with vacuum decay
[$\Lambda$(t)CDM]. In reality,  $\Lambda$(t)CDM cosmologies
constitute the special case in  which the ratio of the dark energy
pressure to its energy density, $\omega$, is exactly $-1$
\cite{Lambda(t),lambdat}. This kind of model may be based on the
idea that dark energy is the manifestation of vacuum quantum
fluctuations in the curved space-time, after a renormalization in
which the divergent vacuum contribution in the flat space-time is
subtracted. The resulting effective vacuum energy density will
depend on the space-time curvature, decaying from high initial
values to smaller ones as the universe expands \cite{1}. As a result
of conservation of total energy, implied by Bianchi identities, the
variation of vacuum density leads either to particle production or
to an increasing in the mass of dark matter particles, two general
features of decaying vacuum or, more generally, of interacting dark
energy models \cite{jsa}.

Naturally, the precise law of vacuum density variation depends on a
suitable  derivation of the vacuum contribution in the curved
background, which is in general a difficult task. In this regard,
however, a viable possibility is initially to consider a de Sitter
space-time and estimate the renormalized vacuum contribution with
help of thermodynamic reasonings as, e.g., those in line with the
holographic conjecture \cite{Holography}. The resulting ansatz is
$\Lambda \approx (H + m)^4 - m^4$, where $H$ is the Hubble parameter
and $m$ is a cutoff imposed to regularize the vacuum contribution in
flat space-time (the next step is to consider a quasi-de Sitter
background, with a slowly decreasing $H$. In this case, the above
ansatz may be considered a good approximation, but with the vacuum
density decaying with time.). In the early-time limit, with $H >>
m$, we have $\Lambda \approx H^4$. By using this scaling law for the
vacuum density and introducing a relativistic matter component, some
of us have obtained from the Einstein equations an interesting
solution with the following features \cite{2}. Firstly, the Universe
undergoes an empty, quasi-de Sitter phase, with $H \approx 1$, which
in the asymptotic limit of infinite past tends to de Sitter solution
with $H = 1$ (in Planck units). However, at a given time,  the
vacuum undertakes a fast phase transition, with $\Lambda$ decreasing
to nearly zero in a few Planck times, producing a considerable
amount of radiation. In this sense, this can be understood as a
non-singular inflationary scenario, with a semi-eternal quasi-de
Sitter phase originating a radiation dominated universe.

The subsequent evolution of vacuum energy essentially depends on the
masses of the produced particles.  If we apply the energy-time
uncertainty relation to  the process of matter production, we will
conclude that massive particles can be produced only at very early
times, when $H$ is very high. Therefore, baryonic particles and
massive dark matter (as supersymmetric particles and axions) stopped
being produced before the time of electroweak phase transition.  On
the other hand, the late-time production of photons and massless
neutrinos must be forbidden by some selection rule, otherwise the
Universe would be completely different from the one observed today.
Thus, if no other particle is produced, the vacuum density stops
decaying at very early times, and for late times we have a standard
universe, with the presence of a genuine cosmological constant (some
thermodynamic considerations, in line with the holographic
principle, permit to infer the value of this constant, leading to
$\Lambda \approx m^6$ \cite{1,2}).

In order to have a decaying vacuum density at the present time, the
produced dark particles should have a mass as small as
$10^{-65}$g\footnote{Some authors associate this mass to the
graviton in a de Sitter background with $H$ and $\Lambda$ of the
orders currently observed \cite{Graviton}.}. Here, we have
considered the possibility of a late-time decaying $\Lambda$, and
compared the consequent cosmological scenario with the constraints
imposed by current observations \cite{3,4}. In such a limit,  $H <<
m$ and $\Lambda \approx m^3 H$, leading again to $\Lambda \approx
m^6$ in the de Sitter limit. From a qualitative point of view, we
have found no important difference between this $\Lambda(t)$
scenario and the flat $\Lambda$CDM model \cite{3}. After the phase
transition described above the Universe enters a radiation-dominated
phase, followed by a matter-dominated era long enough for structure
formation, which tends asymptotically to a de Sitter universe with
vacuum dominating again. The only important novelty, related to
matter production, is a late-time suppression of the density
contrast of matter, which may constitute a potential solution to the
cosmic coincidence problem \cite{5}. Moreover, the analysis of the
redshift-distance relation for supernovae of type Ia, particularly
with the Supernova Legacy Survey (SNLS) data set, has shown a good
fit, with present values of $H$ and the relative density of matter
in accordance with other observations \cite{4}.

In this paper, we go a little further in our investigation and study
new observational consequences of the $\Lambda$(t)CDM scenario
described above. We use to this end distance measurements from type
Ia supernovae (SNe Ia), measurements of the  baryonic acoustic
oscillations (BAO) and the position of the first peak of the cosmic
microwave background (CMB). We show that, besides the interesting
cosmic history of this class of $\Lambda$(t)CDM models, a
conventional, spatially flat $\Lambda$CDM model is only slightly
favored over them by the current observational data.

\section{The model}

In a spatially flat FLRW space-time the Friedmann and the conservation equations can be
written, respectively, as\footnote{We work in units where $M_{\rm{Planck}} = (8\pi G)^{-1/2} = c = 1$.}
\begin{equation} \label{friedmann}
\rho_T = 3H^2\;,
\end{equation}
 and
\begin{equation}
\dot{\rho}_T + 3H(\rho_T + p_T) = 0\;,
\end{equation}
where $\rho_T$ and $p_T$ are the total energy density and pressure,
respectively.  If we consider that  the cosmic fluid is composed of
matter with energy density  $\rho_m$ and pressure $p_m$, and of a
time-dependent vacuum term with energy density $\rho_{\Lambda} =
\Lambda$ and pressure $p_{\Lambda} = - \Lambda$, we obtain
\begin{equation} \label{continuidade}
\dot{\rho}_m + 3H(\rho_m + p_m) = -\dot{\Lambda}\;,
\end{equation}
which shows that matter is not independently conserved, with the decaying vacuum
playing the role of a matter source. Throughout our analysis we assume that baryons are
independently conserved at late times, being not produced at the expenses of the decaying vacuum.
This amounts to saying that we will postulate, in addition to Eq. (\ref{continuidade}), a
conservation equation for baryons, i.e., $\dot{\rho}_b + 3H(\rho_b + p_b) = 0$, where
$\rho_b$ and $p_b$ refer to baryon density and pressure.

From the above equations and considering our late-time ansatz
$\Lambda = \sigma H$ ($\sigma$ is a positive constant of the order
of $m^3$), the evolution equation reads
\begin{equation} \label{evolu�o}
2\dot{H} + 3H^2 - \sigma H = 0\;.
\end{equation}

Now, with the conditions $H > 0$ and $\rho_m > 0$, the  integration
of the above equation leads to the following expression for the
scale factor,
\begin{equation} \label{a}
a(t) = C \left[\exp\left(\sigma t/2\right) - 1\right]^{\frac{2}{3}}\;,
\end{equation}
where $C$ is the first integration constant and the  second one was
chosen so that $a = 0$ at $t = 0$. From these equations, it is
straightforward to show that at early times the above expression
reduces to the Einstein-de Sitter solution whereas at late times it
tends to the de Sitter universe.

As stated earlier, $\Lambda = \sigma H$ and, therefore, $\rho_m =
3H^2 - \sigma H$.  By using solution (\ref{a}), one can also show
that
\begin{equation}\label{rhodust}
\rho_m = \frac{\sigma^2 C^3}{3a^3} + \frac{\sigma^2
C^{3/2}}{3a^{3/2}}\;,
\end{equation}
and
\begin{equation}\label{Lambdadust}
\Lambda = \frac{\sigma^2}{3} + \frac{\sigma^2 C^{3/2}}{3a^{3/2}}.
\end{equation}

The above expressions can be easily understood as follows. The first
terms in the r.h.s. are the expected scaling laws for matter density
and the cosmological constant in the case of a non-decaying vacuum
while the second ones are related to the time variation of the
vacuum density and the concomitant matter production. As expected,
at early times matter dominates, with its density scaling with
$a^{-3}$, and the matter production process is negligible. On the
other hand, at late times the vacuum term dominates, as should be in
a de Sitter universe.

From Eqs. (\ref{friedmann}), (\ref{rhodust}) and (\ref{Lambdadust}),
the evolution of the Hubble parameter as a function of the redshift
can be written as
\begin{equation} \label{Hz}
H(z) = H_0 \left[1 - \Omega_{m} +\Omega_{m} (1 + z)^{3/2}\right],
\end{equation}
where $H_0$ and $\Omega_m$ are, respectively, the present  values of
the Hubble parameter and of the relative energy density of matter.
Note that, due to the matter production,  this expression is rather
different from that found in the context of the standard
$\Lambda$CDM case. In particular, if $\Lambda = 0$ and $\Omega_m =
1$, we obtain $H(z) = H_0 (1 + z)^{3/2}$, leading to $\rho_m =
3H_0^2 (1+z)^3$, as expected for the Einstein-de Sitter scenario.

\section{Observational constraints}

Extending and updating previous results \cite{4}, we  study in this
Section some observational consequences of the class of
$\Lambda$(t)CDM scenarios discussed above. Note that, similarly to
the standard $\Lambda$CDM case, this class of models has only two
independent parameters, $H_0$ and $\Omega_m$ [see Eq. (\ref{Hz})].
The best-fit  values for these quantities will be determined on the
basis of a statistical analysis of recent type Ia supernovae (SNe
Ia) measurements, as given by  the SNLS Collaboration \cite{Legacy},
the distance to the baryonic acoustic oscillations (BAO) from the
Sloan Digital Sky Survey (SDSS) \cite{Eisenstein}, and the  position
of the first peak in the spectrum of anisotropies of CMB radiation
from the three-year Wilkinson Microwave Anisotropy Probe (WMAP)
\cite{WMAP} (for more  details on the statistical analysis discussed
below we refer the reader to Ref.~\cite{refs}).

\subsection{SNe Ia observations}

The predicted distance modulus for a supernova at redshift $z$, given a set of
parameters $\mathbf{s}$, is
\begin{equation} \label{dm}
\mu_p(z|\mathbf{s}) = m - M = 5\,\mbox{log} d_L + 25,
\end{equation}
where $m$ and $M$ are, respectively, the apparent and  absolute
magnitudes, the complete set of parameters is $\mathbf{s} \equiv
(H_0, \Omega_{m})$, and $d_L$ stands for the luminosity distance (in
units of megaparsecs),
\begin{equation}
d_L = c(1 + z)\int_{x'}^{1} {dx
\over x^{2}{H}(x;\mathbf{s})},
\end{equation}
with $x' = {a(t) \over a_0} = (1 + z)^{-1}$ being a  convenient
integration variable, and ${H}(x; \mathbf{s})$ the expression given
by Eq. (\ref{Hz}).

We estimated the best fit to the set of parameters $\mathbf{s}$ by
using a $\chi^{2}$ statistics, with
\begin{equation}
\chi^{2} = \sum_{i=1}^{N}{\frac{\left[\mu_p^{i}(z|\mathbf{s}) -
\mu_o^{i}(z)\right]^{2}}{\sigma_i^{2}}},
\end{equation}
where $\mu_p^{i}(z|\mathbf{s})$ is given by Eq. (\ref{dm}),
$\mu_o^{i}(z)$ is the extinction corrected distance modulus for a
given SNe Ia at $z_i$, and $\sigma^2_i = \sigma^2_{(\mu_{B})} +
\sigma^2_{int}$, where $\sigma^2_{(\mu_{B})}$ is the variance in the
individual observations and  $\sigma^2_{int}$ stands for the
intrinsic dispersion for each SNe Ia absolute magnitude. Since we
use in our analysis the SNLS collaboration sample \cite{Legacy},  $N
= 115$. As discussed in Ref. \cite{4}, the best-fit values for this
analysis is obtained for $h = 0.70$ and $\Omega_m = 0.33$, with
reduced $\chi^2_r = 1.01$. At $95\%$ of confidence level, we also
find $0.69 < h < 0.71$ and $0.28 < \Omega_m < 0.37$. The
$2$-dimensional contours shown in the Figures are obtained from the
traditional frequentist confidence intervals (based on the
$\Delta\chi^2$ approach and assuming that errors are normally
distributed).

\subsection{Baryonic acoustic oscillations}

\begin{figure*}
\vspace{.2in}
\centerline{\psfig{figure=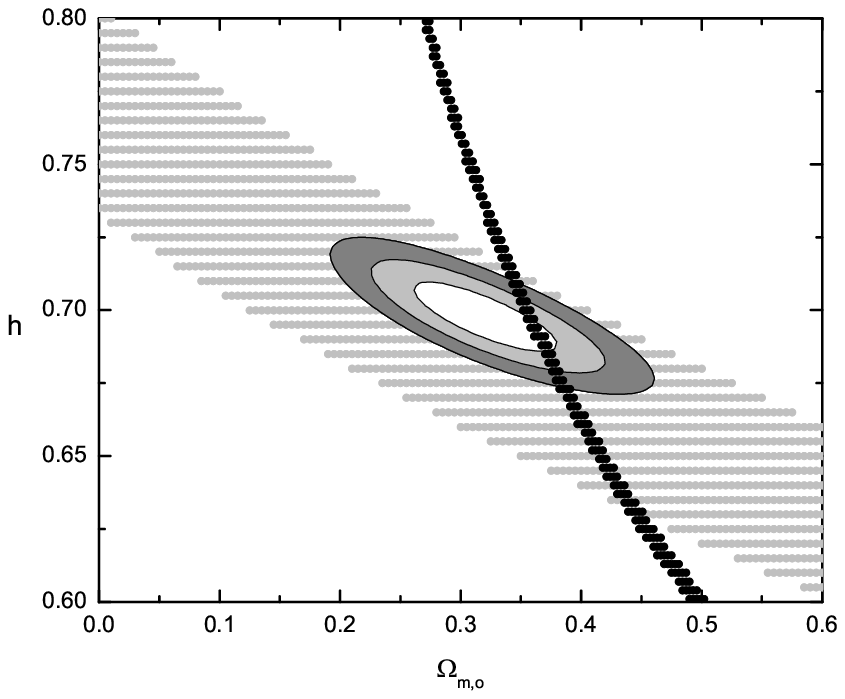,width=3.2truein,height=3.0truein}
\hskip 0.2in
\psfig{figure=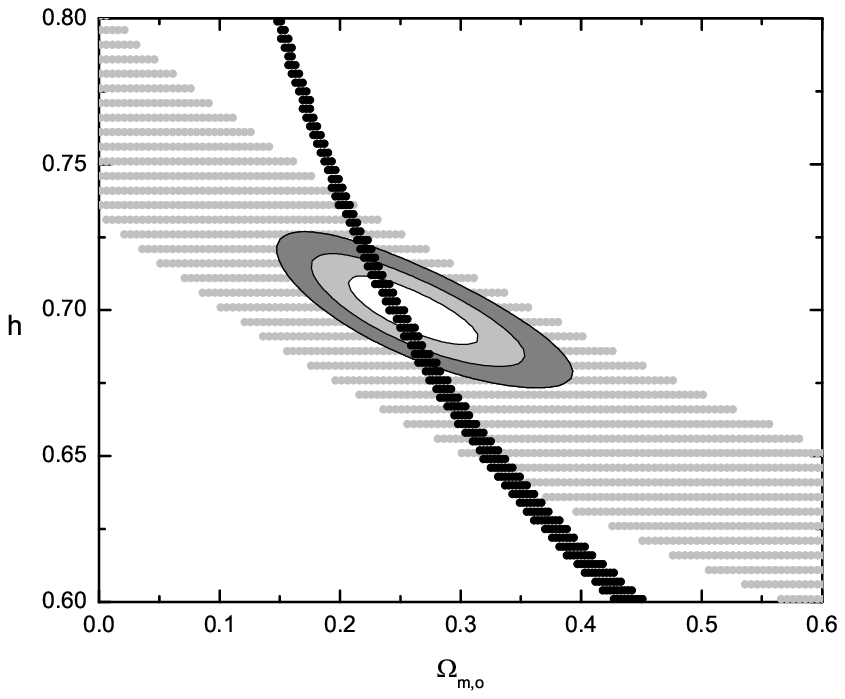,width=3.2truein,height=3.0truein} \hskip
0.1in} \caption{{\bf{Left)}} Superposition of the confidence regions
in  the $\Omega_{\rm{m}} - h$ plane for our analysis of SNe Ia, BAO
and CMB. {\bf{Right)}} The same for a spatially flat $\Lambda$CDM
model.}
\end{figure*}

The use of BAO to test dark energy models is usually made by means
of the parameter ${\cal{A}}$, i.e.,  \cite{Eisenstein}
\begin{equation}
{\cal{A}} = D_V \frac{\sqrt{\Omega_m H_0^2}}{zc}\;,
\end{equation}
where $z = 0.35$ is the typical redshift of the sample and $D_V$ is
the dilation scale, defined as
\begin{equation} \label{DV}
D_V(z) = \left[D_M(z)^2 \frac{zc}{H(z)}\right]^{1/3}\;,
\end{equation}
with the comoving distance $D_M$ given by
\begin{equation}\label{DM}
D_M(z) = c\int_0^z \frac{dz'}{H(z')}\;.
\end{equation}

An important aspect worth emphasizing at this point is  that the
value of ${\cal{A}}$ is obtained from the data in the context of a
$\Lambda$CDM model, and can be considered a good approximation only
for some class of dark energy models \cite{EarlyDE}. In particular,
two conditions are implicitly assumed to be valid \cite{Eisenstein}.
First, the evolution of matter density perturbations during the
matter-dominated era must be similar to the $\Lambda$CDM case, at
least until the characteristic redshift $z = 0.35$. Second, the
comoving distance to the horizon at the time of equilibrium between
matter and radiation must scale with $(\Omega_m H_0^2)^{-1}$.
However, none of the above conditions are satisfied in the present
model because of the matter production associated to the vacuum
decay. As will be shown in a forthcoming publication \cite{5}, if
matter is homogeneously produced there is a suppression in its
density contrast for $z < 5$, that is, after the period of galaxy
formation (this will eventually imply a higher value of $\Omega_m$
in order to fit the observed mass power spectrum).

On the other hand, as radiation is independently  conserved, its
relative energy density for $z >> 1$ is given by $\Omega_r z^4$,
where $\Omega_r$ is its present value. With the help of (\ref{Hz})
we can see that, in the same limit $z >> 1$, the relative density of
matter is $\Omega_m^2 z^3$ (with the extra factor $\Omega_m$ being
due to the matter production between $t(z)$ and the present time).
By equating the two densities, we obtain the redshift of equilibrium
between matter and radiation, given by $z_{rm} =
\Omega_m^2/\Omega_r$. Therefore, after including conserved radiation
into (\ref{Hz}) (see equation (\ref{Hgeral}) below) and taking $z >>
1$, we have $r_H(z_{rm}) \approx c\sqrt{\rho_r/6}(\Omega_m
H_0)^{-2}$, where $\rho_r$ is the present value of the radiation
density (while in the $\Lambda$CDM case we would obtain $r_H(z_{rm})
\approx c\sqrt{\rho_r/6}(\Omega_m H_0^2)^{-1}$, as stated above).
Thus, one can see that the parameter ${\cal{A}}$ is not appropriate
to test the model, and we will use instead the dilation scale $D_V$,
which is weakly sensitive to the cosmological evolution before $z =
0.35$. By combining our function $H(z)$, given by (\ref{Hz}), into
(\ref{DV})-(\ref{DM}) we can find the region in the $\Omega_m-H_0$
plane which gives the observed value $D_V = (1370 \pm 64)$ Mpc
($1\sigma$) \cite{Eisenstein}. The BAO bands in the $\Omega_m-H_0$
parametric space are shown in Fig. 1.

\subsection{The first peak of CMB}

\begin{figure*}
\vspace{.2in}
\centerline{\psfig{figure=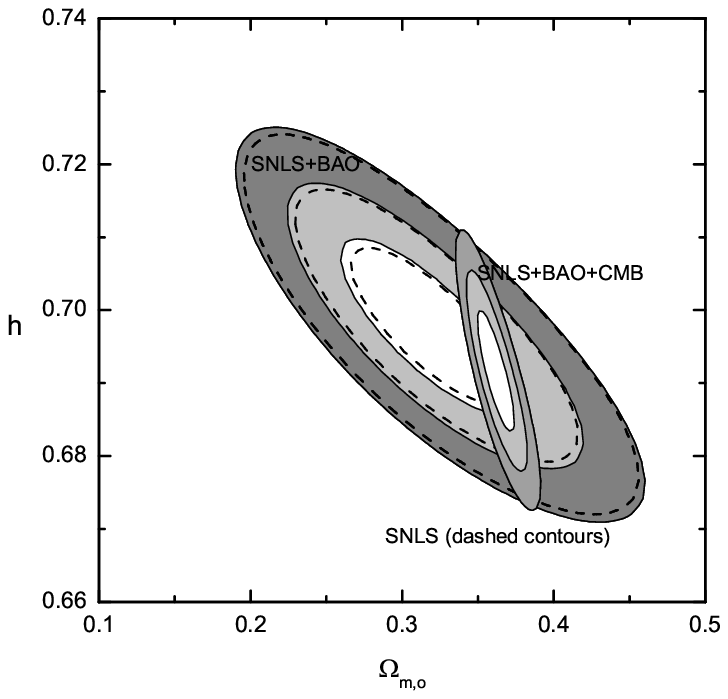,width=3.7truein,height=3.2truein}
\psfig{figure=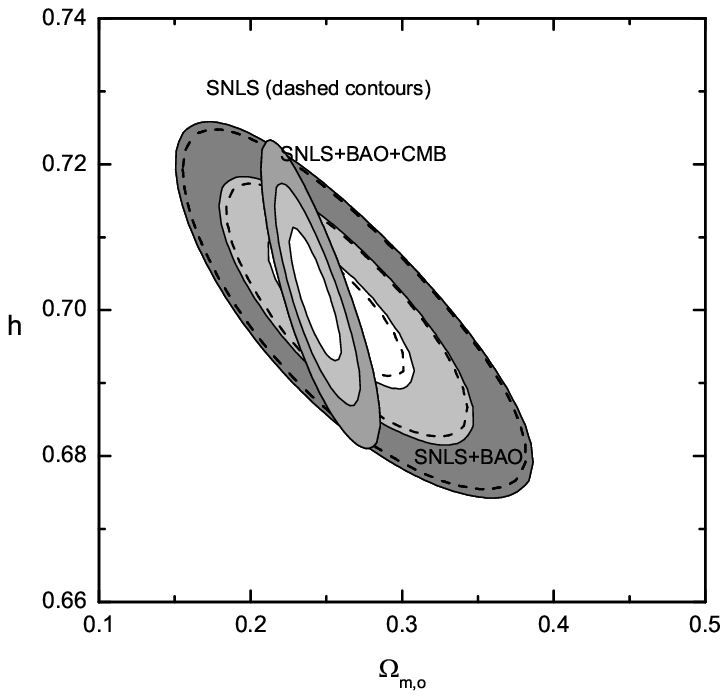,width=3.7truein,height=3.2truein} \hskip
0.1in} \caption{{\bf{Left)}} Confidence regions (68.3\%, 95.4\% and
99.7\%) in the $\Omega_{\rm{m}} - h$ plane for a joint analysis
involving SNe Ia, BAO and CMB data. As indicated, the dashed lines
stand for the SNe Ia results of Ref. \cite{4}. In both panels, the
largest contours represent the joint analysis of SNe Ia plus BAO
measurements whereas the smallest ones arises when the CMB data are
included in the analysis. {\bf{Right)}} The same for a spatially
flat $\Lambda$CDM model.}
\end{figure*}

The two tests we have previously described depend on the  physics at
low and intermediate redshifts (until $z \sim 1$) and will lead, as
we will see, to very similar results. As a complementary test,
involving high-$z$ measurements, let us consider the position of the
first peak in the spectrum of CMB anisotropies. Since in the present
model there is no production of baryonic matter or radiation and the
spatial curvature is null, we expect that a correct position of the
first peak is enough to guarantee a spectral profile similar to the
$\Lambda$CDM one, provided the spectrum of primordial fluctuations
is the same.

In the context of a large class of dark energy models this test is
performed by comparing the predicted shift parameter with the
$\Lambda$CDM value. However, this is only valid if the acoustic
horizon at the time of last scattering is the same \cite{Raul}. This
is not true in the present model because, as we have already shown,
for the same values of $H_0$ and $\Omega_m$ we have different
expressions for cosmological parameters at high redshifts, due to
the process of matter production. Therefore, in order to perform a
correct test, we have to explicitly calculate the acoustic scale in
the model and then compare with the measured position of the peak.

The acoustic scale, defined as the ratio between the comoving
distance to the surface of last scattering and the radius of the
acoustic horizon at that time, is then given by
\begin{equation}\label{lA}
l_A = \frac{\pi \int_0^{z_{ls}}
\frac{dz}{H(z)}}{\int_{z_{ls}}^{\infty}\frac{c_s}{c}\frac{dz}{H(z)}},
\end{equation}
where $z_{ls}$ is the redshift of last scattering, and
\begin{equation}\label{cs}
c_s = c \left( 3 + \frac{9}{4}\frac{\Omega_b}{\Omega_{\gamma}z}
\right)^{-1/2}
\end{equation}
is the sound velocity. Here, $\Omega_b$ and $\Omega_{\gamma}$
are, respectively, the present relative energy densities of baryons and
photons.

The function $H(z)$ to be used in Eq. (\ref{lA}) must now include
radiation. As this component is independently conserved, scaling
with $a^{-4}$, the appropriate generalization of Eq. (\ref{Hz}) is
given by\footnote{The inclusion of a conserved component of
radiation changes the dynamics and, consequently, the evolution of
$\Lambda(z)$ and $\rho_m(z)$. Thus, rigorously speaking the
generalization of $H(z)$ would require a reanalysis of the dynamics.
Nevertheless, as $\Omega_r \approx 10^{-4}$, when the decaying
vacuum and  matter production begin to be important, the radiation
term is negligible. For this reason, Eq. (\ref{Hgeral}) can be
considered a good approximation. Indeed, a numerical analysis in the
range $0 < z < 10^4$ showed a difference between Eq. (\ref{Hgeral})
and the exact $H(z)$ as small as $0.01\%$.}
\begin{equation}\label{Hgeral}
\frac{H(z)}{H_0} \approx \left\{ \left[1 - \Omega_{m} +\Omega_{m} (1
+ z)^{3/2}\right]^2 + \Omega_r (1+z)^4 \right\}^{1/2}.
\end{equation}

Therefore, apart from our two free parameters $H_0$  and $\Omega_m$,
in order to determine the acoustic scale we need the present values
of the energy densities of radiation, photons and baryons, as well
as an expression for $z_{ls}$. Since radiation and baryons are
independently conserved, and we want to preserve the spectrum
profile as well as the nucleosynthesis constraints, we will take for
these densities the same values obtained from CMB observations in
the context of the $\Lambda$CDM model \cite{CMB}: $\Omega_{\gamma}
h^2 \approx 2.45 \times 10^{-5}$, $\Omega_r h^2 \approx 4.1\times
10^{-5}$ and $\Omega_b h^2 \approx 0.02$.

Concerning $z_{ls}$, its value is not in principle the  same as in
the $\Lambda$CDM case. To determine its value for a given pair
($\Omega_m$,$H_0$) we will proceed as follows. First, we obtain the
redshift of last scattering $z^*_{ls}$ in the $\Lambda$CDM model by
means of the current fitting formula \cite{Raul}, and then we impose
that the optical depth has to be the same in both models, that is,
\begin{equation}
\int_0^{z_{ls}}\frac{\Gamma(z)}{H(z)}\frac{dz}{1+z} = \int_0^{z^*_{ls}}
\frac{\Gamma(z)}{H^*(z)}\frac{dz}{1+z},
\end{equation}
where $\Gamma$ is the rate of photon scattering and $H^*$ is the
Hubble parameter in the $\Lambda$CDM context. For the current
intervals of $H_0$ and $\Omega_m$, the relative differences between
$z_{ls}$ and $z^*_{ls}$ are typically as small as $1\%$.

For a scale invariant $\Lambda$CDM model with spectral index $n =
1$,  the position of the first peak after including the effect of
plasma driving is given by the fitting expression \cite{Tegmark}
\begin{equation}\label{l1}
l_1 = l_A (1 - \delta_1),
\end{equation}
where
\begin{equation}
\delta_1 = 0.267 \left( \frac{r}{0.3} \right)^{0.1},
\end{equation}
with $r \equiv \rho_{\gamma}(z_{ls})/\rho_m(z_{ls})$. Since the
plasma driving depends essentially on pre-recombination physics, we
will therefore assume that the above fitting formulae are a good
approximation to the present $\Lambda(t)$ model. In our case, the
parameter $r$ is given by
\begin{equation}
r = \frac{\Omega_{\gamma}}{\Omega_m^2} z_{ls}.
\end{equation}

Finally, from the above results we can determine the region of  the
$\Omega_m$-$H_0$ plane for which the first peak has the position
currently observed by WMAP, i.e.,  $l_1 = 220.8 \pm 0.7$ ($1\sigma$)
\cite{WMAP}. As in the case of BAO, the CMB  bands in the
$\Omega_m-H_0$ parametric space is shown in Fig. 1.

\subsection{Results and discussions}

The superposition of the confidence regions and bands corresponding
to our analysis of SNe Ia, BAO and CMB observations is shown in
Figure 1. For the sake of comparison, the right panel shows the same
analysis for the standard $\Lambda$CDM model.  From these analysis,
we note that, although not very restrictive and parallel to the SNe
Ia contours, the BAO bands can be statistically combined with SNe Ia
data providing constraints on the $\Omega_m - H_0$ plane. At
$2\sigma$ level, we find $h = 0.70 \pm 0.01$ and $\Omega_m = 0.32\pm
0.05$, with reduced $\chi^2_r \simeq 1.00$. The inclusion of CMB
data into our analysis in turn makes the complete joint analysis
more restrictive but also shows that the concordance is as good as
in the standard $\Lambda$CDM case, and that the $\Lambda$(t)CDM
model discussed here cannot be ruled out. It can be anticipated from
Figure 1 that the data will prefer higher values of the matter
density parameter, while the Hubble parameter is expected to be
slightly smaller than the current accepted values. In order to
confirm this qualitative discussion, Figure 2 shows the results of
our joint statistical analysis (SNe Ia + BAO + CMB). At $2\sigma$
level, we find $h = 0.69 \pm 0.01$ and $\Omega_m = 0.36 \pm 0.01$,
with reduced $\chi^2_r = 1.01$. For the sake of comparison, the same
analysis for the $\Lambda$CDM case is also shown in the right panel
of Fig. 2.

Finally, with the above best-fit values for the parameters,  we can
calculate the total expanding age for this class of $\Lambda$(t)CDM
models, given by  \cite{3,4}
\begin{equation} \label{age}
t_0 = H_0^{-1}\frac{\frac{2}{3}\ln(\Omega_{\rm{m}})}{\Omega_{\rm{m}}
- 1} \simeq 15.0 \hspace {0.03cm} \mbox{Gyr}\;,
\end{equation}
as well as the redshift of transition from a decelerated  expansion
to the current accelerating phase, i.e.,  $z_T \approx 1.3$. Note
that both values are slightly higher than, but of the same order of,
the standard ones.

\section{Final remarks}

By using the most recent cosmological observations,  we have
discussed  the observational viability of a class of $\Lambda$(t)CDM
scenarios in which $\Lambda \propto H$, as well as a possible way to
distinguish it from the standard $\Lambda$CDM model in what concerns
the general characteristics of the predicted cosmic evolution. As
discussed earlier, these $\Lambda$(t)CDM models have some
interesting features as, e.g., the association of dark energy with
vacuum fluctuations, the circumvention of the cosmological constant
problem by subtracting the flat space-time contribution from the
curved space-time vacuum density, and the possible (but not
necessary) link between dark matter and massive gravitons.

We have presented some quantitative results which clearly show that,
even in the current stage of the Universe evolution, our decaying
vacuum scenario is very similar to the standard one. We have
statistically tested the viability of the model by using recent SNe
Ia observations and measurements of the BAO and the first peak of
the CMB spectrum. At 95.4\% c.l., a joint analysis involving SNe Ia
+ BAO provides the intervals $0.69 \leq h \leq 0.71$ and $0.28 \leq
\Omega_m \leq 0.37$, which are in good agreement with the values of
the Hubble and the matter density parameters obtained from
independent analysis \cite{hst,omega}. When the position of the
first peak of CMB anisotropies is included in the analysis, the
best-fit value for the relative matter density is increased,
$\Omega_m \simeq 0.36$. This result cannot rule out the model, and
it may be indicating that, besides the interesting cosmic history of
this class of $\Lambda$(t)CDM models, a conventional, spatially flat
$\Lambda$CDM model is only slightly favored over them by the current
observational data. Still on the best-fit for $\Omega_m$, we note
that such a higher matter density is something to be more
investigated by means of other cosmological or dynamical tests as,
e.g., the predicted mass power spectrum in the context of the model.
As we have discussed earlier, a higher matter density is necessary
to compensate the late-time suppression of the density contrast
owing to matter production \cite{5}. Another possibility will be
provided by future supernovae observations, since the present model
starts to diverge from the standard one for higher redshifts
\cite{4}.

Finally, we should also emphasize two aspects to be considered
before a definite conclusion about the comparison between the
present model and the flat standard scenario. First, that in our
study of CMB we have used parameter values and fitting formulae that
are strictly correct for the $\Lambda$CDM case, particularly the
expression giving the position of the first peak for a given
acoustic scale [equation (\ref{l1})]. In spite of our reasons to
consider that use as a good approximation, it can lead to bias in
our results. If that is the case, only a more complete analysis of
CMB would rule out or corroborate the model. The second aspect is of
theoretical character. As discussed in the Introduction, our ansatz
for the variation of $\Lambda$, if genuine, is a good approximation
only for quasi-de Sitter backgrounds. Naturally, our Universe,
although dominated by a cosmological term, is far away from the
asymptotic de Sitter state, with matter still giving an important
contribution to the cosmic fluid. Only a more profound theoretical
study, on the basis of quantum field theories in the expanding
background, could establish the degree of applicability and limits
of that approximation.

\section*{Acknowledgements}

The authors are grateful to Deepak Jain, Max Tegmark and Raul Abramo
for useful references. This work is partially supported by CNPq and
CAPES. JSA is also supported by FAPERJ No. E-26/171.251/2004.

{}

\end{document}